# Transport in Suspended Monolayer and Bilayer Graphene Under Strain: A New Platform for Material Studies

*Hang Zhang[†][*], Jhao-Wun Huang[†], Jairo Velasco Jr, Kevin Myhro, Matt Maldonado, David Dung Tran, Zeng Zhao, Fenglin Wang, Yongjin Lee, Gang Liu, Wenzhong Bao, and Chun Ning Lau[*]*

Department of Physics and Astronomy, University of California, Riverside, CA 92521

Emails: lau@physics.ucr.edu, hang.zhang@caltech.edu

[†] These authors contribute equally to this work.

**Abstract：**

We develop two types of graphene devices based on nanoelectromechanical systems (NEMS), that allows transport measurement in the presence of *in situ* strain modulation. Different mobility and conductance responses to strain were observed for single layer and bilayer samples. These types of devices can be extended to other 2D membranes such as $MoS_2$, providing transport, optical or other measurements with *in situ* strain.

TOC figure:

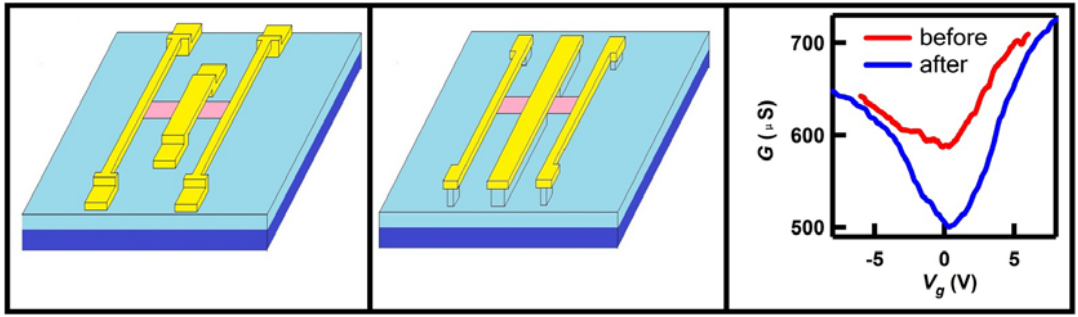

1. Introduction

Graphene is a two-dimensional carbon allotrope. Since its first isolation onto insulating substrates[1] and the subsequent development of wafer scale synthesis technology[2-4], graphene has attracted wide attention as a promising candidate for next generation electronics materials[5-7] [8-13]. As nature's thinnest membrane, graphene's electronic properties are also intimately related to its morphology and/or strain; thus inducing strain may be used to modify the transport properties or band structure of pristine graphene[14-18]. Prior works have demonstrated strain in graphene can be controlled via controlling temperature[17, 19] or chemical modifications[20-23], though *in situ* control of strain was not achieved. In ref. 24, Huang *et al* combined transport studies and *in situ* strain control by loading suspended graphene samples with a nano-tip in the chamber of a scanning electron microscope (SEM), though only marginal changes in electrical properties are observed upon application of ~<1% strain; moreover, exposure to electron beam irradiation degrades sample quality[24-26]. Thus there is still much to be explored in transport studies on ultra-thin graphene films with *in situ* strain control.

In this letter we present transport measurements of suspended monolayer graphene (MLG) and bilayer graphene (BLG) nanoelectromechanical (NEM) devices, who allows *in situ* modification of strain up to 5%. We study the device behavior before and after repeated straining cycles. For MLG devices, the two-terminal conductance $G$ vs. gate voltage $V_g$ curve becomes smoother, and the minimum conductance shows minimal change (<1%), in agreement with prior results[24]. For BLG, the minimum conductance decreases by more than 10% and field effect mobility increases. The different behaviors between MLG and BLG devices may arise from the relative shear between the two layers in BLG, or the presence of stacking domains (e.g. AB-BA) whose boundaries are particularly susceptible to strain. Our results underscore the rich interplay between strain and transport offered by suspended devices. Furthermore, these types of NEMS devices are compatible with optical measurements and can be used to study other two-dimensional materials.

## 2.  Experimental section

2.1 Device fabrication: Graphene sheets were extracted from bulk graphite using standard mechanical exfoliation techniques on top of $SiO_2$/Si substrates or a layer of the LOR resist. The number of layers was initially identified via optical microscopy and subsequently confirmed with Raman spectroscopy after completion of transport measurements (Figure 1a). To perform transport measurement and *in situ* stretching, we fabricated nano-electromechanical system (NEMS)-based graphene devices, using two different techniques:

In Method A, devices were fabricated with multi-level lithography based on the resists consist of LOR layer on top of PMMA layer. Detailed fabrication process is described in our previous work[27] (Figure 1b). Devices thus fabricated have relatively large areas, and graphene are "held" up by electrodes that are suspended above the SiO$_2$/Si substrates (Figure 1c). The central electrode was designed to be wider and shorter than the neighboring electrodes, so that it can sustain higher actuating voltages.

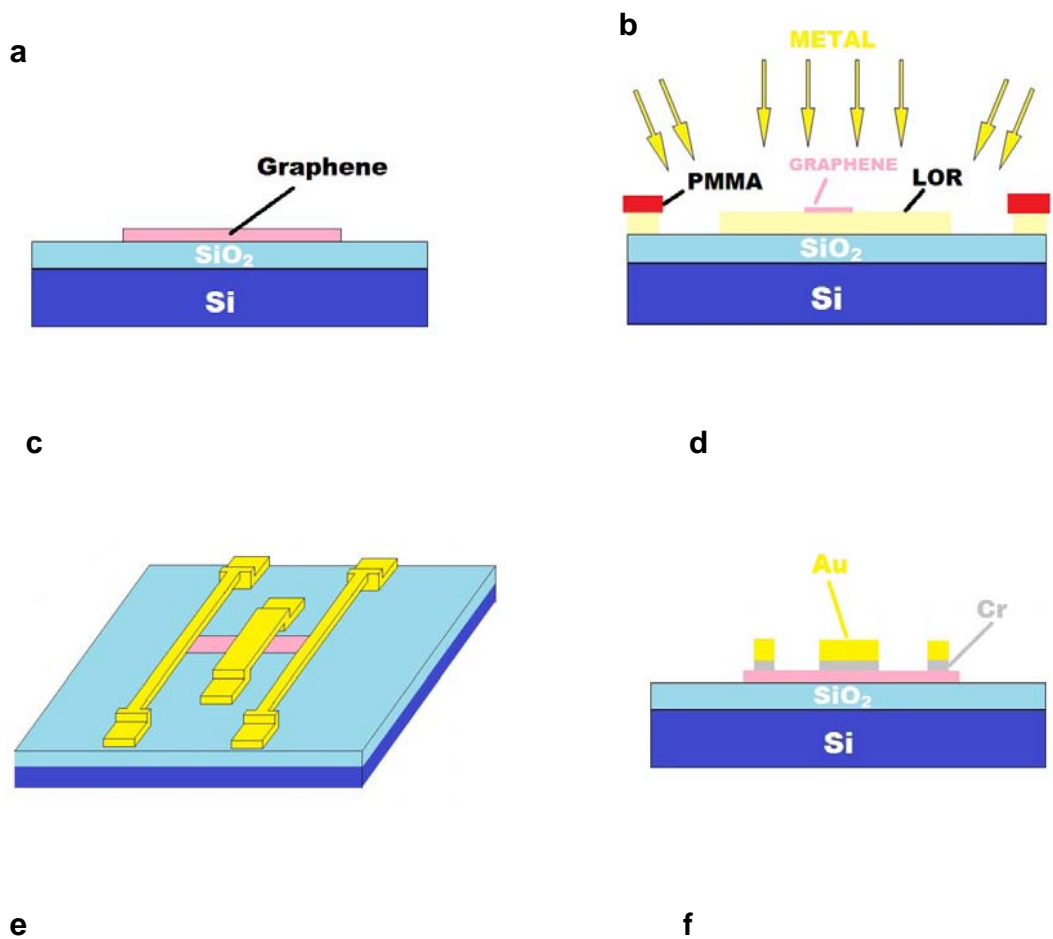

a

b

c

d

e

f

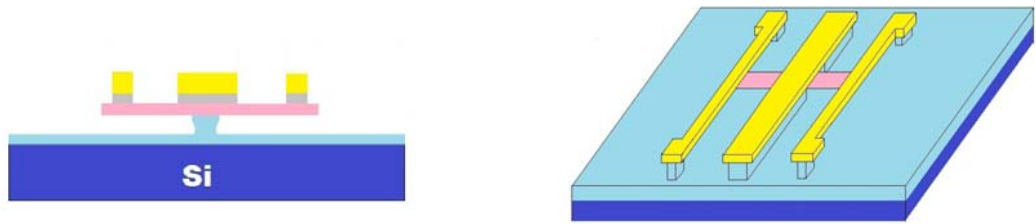

**Figure 1.** (a). Graphene sample exfoliation and identification. (b). Fabrication process using Method A and angled deposition. (c). Three dimensional schematic of a device fabricated with Method A. (d). Fabrication of a device using Method B, which is initially non-suspended. (e) BOE etching selectively removes $SiO_2$ underneath graphene samples and electrodes. (f). Three dimensional schematic of a device fabricated with Method B.

In Method B, which is used to fabricate the majority of the devices, three Cr/Au (10nm/150nm) electrodes were attached to graphene flakes using standard electron beam lithography (Figure 1d). Then the whole device was submerged into buffered oxidant etchant (BOE) solution for 90~120s.[28-30] For each device, the central electrode is designed to be 2-3 times wider (800nm~1000 nm wide, 25 ~40 μm long ) than the two neighboring electrodes (300~400nm, 20 ~30 μm long). All electrodes are anchored by large contact pads at the ends. By controlling etching time, we can control the extent of $SiO_2$ etched underneath the electrodes and graphene flake, so that narrower features and graphene flakes are suspended, whereas the wider features (central electrodes and anchors) remain supported by residual of $SiO_2$ underneath (Figure 1e). After etching, the device was transferred into water and isopropyl alcohol (IPA) in succession, in order to rinse and cover the sample with a liquid with lower surface tension. Finally, the device was taken out from hot IPA (to further decrease surface tension of IPA) and placed onto a hot plate at 70℃. The fabrication process is very robust: despite the fragility of suspended graphene devices, the yield is ~90%. Figure 1f illustrates the schematics of a typical device with this method.

2.2 Transport Measurements of the Devices: The devices were placed in a custom-built helium cryostat. All the measurements were performed in a high vacuum environment. The temperature of the devices was measured with a semiconductor thermometer mounted in close proximity to the chip carriers. Data were acquired by National Instrument PCI-6251 card controlled by a C++ based program.

2.3 SEM In Situ Imaging: To avoid contamination and damage caused by SEM imaging, all SEM characterizations were done on devices after finishing all the transport measurements or "SEM-imaging-only" devices.

## 3. Results and discussion

Figure 2a illustrates the general principle of applying *in situ* strain. A suspended electrode and the back gate (Si substrate) formed a capacitor. Initially both electrodes and the back gate are grounded, thus they remain parallel, as outlined by solid lines in Figure 2a. Upon applying the actuating voltage (bias voltage between electrodes and back gate), the electrostatic force induces deflection in the outer suspended or longer electrodes toward the substrate, whereas the central electrode (that is shorter, wider and/or partially supported by the substrate) remain suspended; thus the far ends of the attached graphene sheet move downward accordingly. From the geometry of our device, we can estimate the strain $\gamma$ exerted on graphene

$$\gamma = \sqrt{1 + \frac{h^2}{L_0^2}} - 1 \qquad (1)$$

where $h$ denotes the maximum vertical deflection of the suspended electrode under the electrostatic force, $L_0$ indicates the initial length of suspended graphene sample. We estimate that at maximum load, up to 5% strain can be induced in the graphene sheets.

Figure 2b-c show a device fabricated using method A at gate voltage $V_g=0$ and 30V, respectively. Initially all electrodes and the graphene sheet are well-suspended. When the gate voltage ramps, the narrower electrode on the left slowly deflects downward;

at $V_g$=30 V, it buckles and collapses to the substrate. (The *in situ* stretching process is shown in the video of supporting information.) This collapse is irreversible even when $V_g$ is reduced to 0. We note that when the measurement is repeated on a control device with the same geometry but without the graphene flake, the suspended electrode collapses at much smaller voltage $V_g$~7V. Since the electrostatic force is proportional to $V_g^2$, we estimate that at $V_g$=30V, ~95% of the electrostatic force is exerted on the graphene sheet. Figure 2d shows another device before and upon applying $V_g$~100V. Periodic ripples appears in the graphene sheet afterwards arises from the longitudinal strain induced[17].

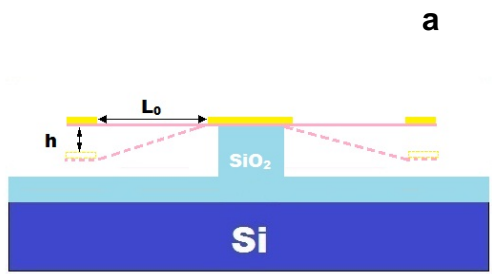
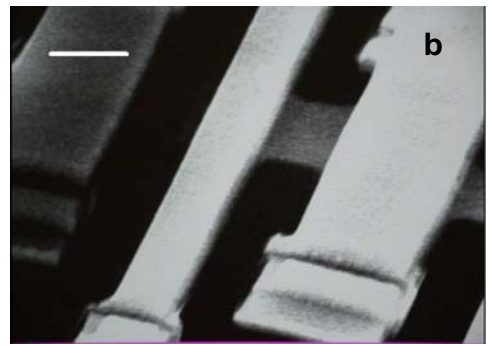

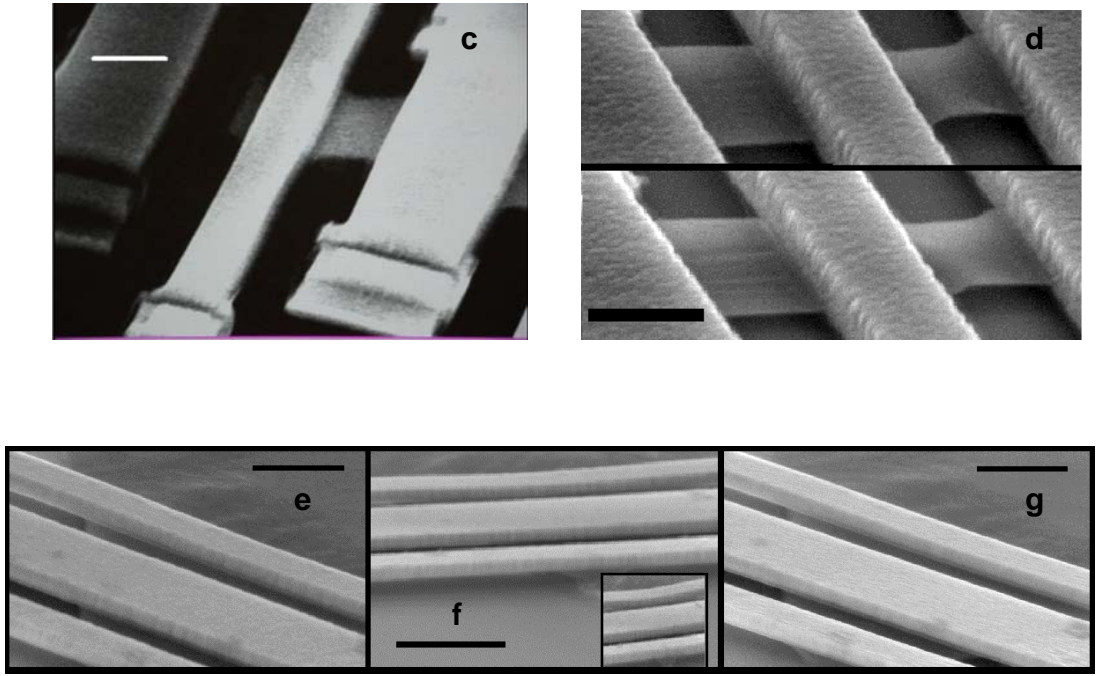

**Figure 2.** (a). Schematic of a device with and without applying the actuating voltage between electrodes and back gate. (b-c) SEM images of a device fabricated with Method A at $V_g=0$ and $V_g=30V$, respectively. Scale bars: 2μm. (d). SEM images of another graphene sample fabricated before (upper panel) and after stretched (lower panel). Scale bars: 2μm. (e,f,g) SEM images of a device made by Method B at $V_g=0$, $V_g=50V$, and when $V_g$ is returned to 0. Scale bars: 1μm in (e) and (g), 2μm in (f). The inset in (f) shows a zoom-in image of the deflected region.

For devices fabricated with method B, the suspended electrodes can reversibly move between parallel and deflected positions. Figure 2e shows a device made by method B before stretching. Figure 2f displays SEM image of the same device is stretching a suspended sample under a $V_g$ ~50V. A zooming in image shows one narrower electrode clearly deflected toward the substrate (inset in f). Figure 2g displays the same device when $V_g$ is returned to 0V, and the suspended electrode returns to its original height. To avoid collapsing the samples, we typically limit the actuating voltage to less than 60V.

To perform transport measurements, the devices are cooled down to 4.2K in vacuum. Current annealing was applied to remove contaminants on the graphene sheet. The devices are first characterized by measuring its conductance $G$ as a function of $V_g$; (Figure 3a, red curve) here $V_g$ is limited to $<\pm 10$V, so that strain is negligible. All devices show repeatable $G(V_g,)$ curves over such small $V_g$ range.

After extracting data from its initial state, we start stretching the sample by gradually ramping up actuating voltage to -50V. Figure 3b shows the conductance changes as time elapsed, when the actuating voltage is maintained at 50V. The conductance fluctuates noticeably and decreased by more than 20 µS (~ 1%). We note that this effect cannot be explained by the changing capacitance between graphene and the gate – at the strained position, the device has stronger coupling to the gate, thus should give rise to a *higher* conductance value. Thus the modulation in conductance must be induced by movement of the electrode itself, *e.g.* strain and/or changing the graphene-electrode interface.

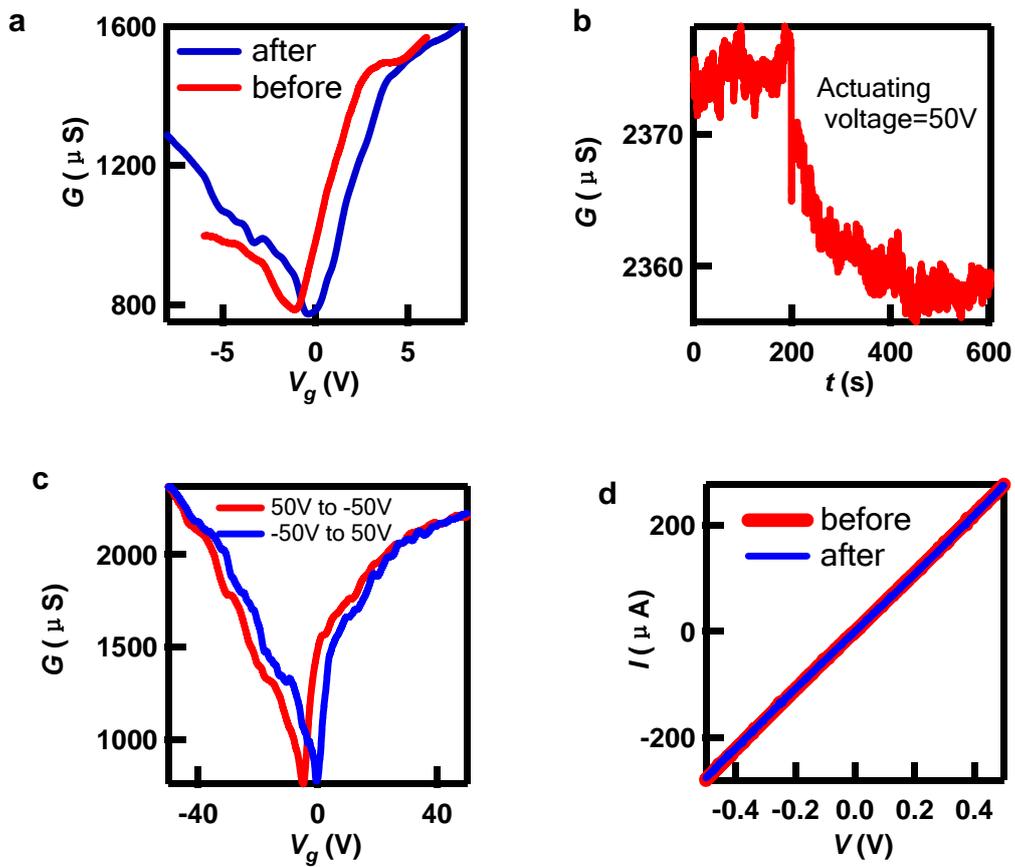

**Figure 3.** (a) Conductance as a function of back gate voltage, before (red curve) and after (blue curve) stretching process, from a single layer graphene device. (b). Conductance *vs.* time when the actuating voltage is kept at 50V. (c). Conductance as a function of back gate after several stretching cycles. (d). *IV* curves of a typical single layer device before and after stretching process.

The tension in the sample is then released by lowering the actuating voltage back to 0V, and characterized again by measuring $G(V_g)$ for limited $V_g$ range (Figure 3a, blue curve). For single layer graphene, minor changes such as slightly improved mobility are observed, but generally the minimum conductivity and the current-voltage *(I-V)* characteristics (Figure 3d) stay relatively constant. After several repeated sweeping cycles (between +/- 50V), the gate response became stable (Figure 3c) even at large gate voltage.

Compared with single layer samples, bilayer devices behaves quite differently. Figure 4a shows the $G(V_g)$ curves before and after stretching from a typical bilayer devices. After releasing from external strain, the curve becomes steeper and smoother, and the mobility improves. Interestingly, the minimum conductance decreased considerably. This can also been seen in the *I-V* curves, which is more non-linear after stretching (Figure 4b). Typically, after stretching process, the conductance of bilayer devices decreases by 10%~15%. After several stretching cycles several times, the device's $G(V_g)$ becomes stable (Figure 4c) with the improved mobility and lower minimum conductance. The device shows no appreciable change in appearance after the stretching cycles (Figure 4d).

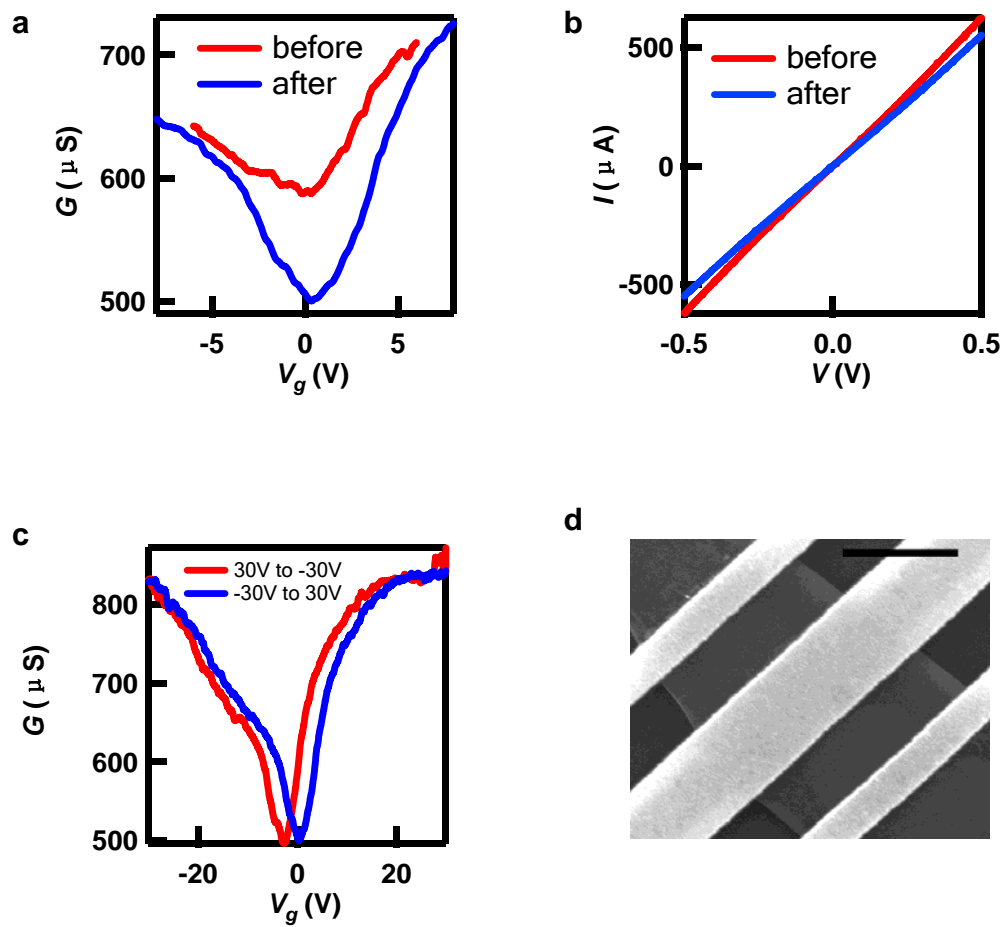

**Figure 4.** (a) Conductance as a function of back gate voltage, before (red curve) and after (blue curve) stretching process, from a bi-layer graphene device. (b). *IV* curves from one typical bi-layer device before and after the stretching process. (c).

Conductance as a function of back gate after several cycles. (d) SEM image of one bilayer graphene device after stretching. Scale bar: 1μm.

These intriguing observations suggest the rich interplay between strain and transport offered by suspended devices. The improvement in device mobility likely arises from releasing the strain or ripples that are built-in during the fabrication process. The different behaviors between single layer and bilayer devices are particularly intriguing, *e.g.* the significant decrease in minimum conductance is unique to bilayer devices. A possible explanation is the improved contact at the electrode-graphene interface; however, one expects that this scenario should occur in single-layer devices as well. We also exclude strain-induced cracks, which should occur at much higher strain[31, 32] and also lead to lower mobility. Our present proposal is that the decrease in minimum conductance may be caused by relative shift and/or shear between two layers induced by the stretching cycles,[33-35] or the presence of AB-BA stacking domains whose boundaries may shift in response to strain.[36, 37] This hypothesis can be verified by low temperature transport measurements, as the modified band structure is expected to lead to reduced density of states and different Landau level spectrum than that of an AB-stacked bilayer graphene.

**4.    Conclusion**

In conclusion, we developed two types of NEMS-like devices to stretch suspended single crystal graphene samples and perform *in situ* measurements. The stretching process can be observed via SEM imaging. Transport property investigation shows that after stretching process, the gate response of conductance from graphene samples improved, and dramatic decrease in minimum conductance is observed in bi-layer graphene samples. The experimental system and method introduced in this work provides a new approach in strain engineering researches.

**Supporting Information**

Supporting Information is available from the Elsevier website or from the authors.

**Acknowledgements**

We acknowledge the support by NSF CAREER DMR/0748910, NSF/1106358, the FAME Center, UC Lab Fees Program, DMEA/H94003-10-2-1003 and ONR.

**References**

[1]     Novoselov KS, Geim AK, Morozov SV, Jiang D, Zhang Y, Dubonos SV, et al. Electric field effect in atomically thin carbon films. Science. 2004;306(5696):666-9.
[2]     Berger C, Song Z, Li X, Wu X, Brown N, Naud C, et al. Electronic confinement and coherence in patterned epitaxial graphene. Science. 2006;312(5777):1191-6.
[3]     Li X, Cai W, An J, Kim S, Nah J, Yang D, et al. Large-area synthesis of high-quality and uniform graphene films on copper foils. Science. 2009;324(5932):1312-4.

[4]     Sun Z, Yan Z, Yao J, Beitler E, Zhu Y, Tour JM. Growth of graphene from solid carbon sources. Nature. 2010;468(7323):549-52.
[5]     Chen JH, Jang C, Xiao SD, Ishigami M, Fuhrer MS. Intrinsic and extrinsic performance limits of graphene devices on SiO2. Nat Nanotechnol. 2008;3(4):206-9.
[6]     Du X, Skachko I, Barker A, Andrei EY. Approaching ballistic transport in suspended graphene. Nat Nanotechnol. 2008;3(8):491-5.
[7]     Bao W, Zhao Z, Zhang H, Liu G, Kratz P, Jing L, et al. Magnetoconductance oscillations and evidence for fractional quantum Hall states in suspended bilayer and trilayer graphene. Physical review letters. 2010;105(24):246601.
[8]     Geim AK, Novoselov KS. The rise of graphene. Nat Mater. 2007;6(3):183-91.
[9]     Novoselov KS, Geim AK, Morozov SV, Jiang D, Katsnelson MI, Grigorieva IV, et al. Two-dimensional gas of massless Dirac fermions in graphene. Nature. 2005;438(7065):197-200.
[10]    Zhang Y, Tan YW, Stormer HL, Kim P. Experimental observation of the quantum Hall effect and Berry's phase in graphene. Nature. 2005;438(7065):201-4.
[11]    Fuhrer MS. Graphene: Ribbons piece-by-piece. Nat Mater. 2010;9(8):611-2.
[12]    Lin YM, Dimitrakopoulos C, Jenkins KA, Farmer DB, Chiu HY, Grill A, et al. 100-GHz Transistors from Wafer-Scale Epitaxial Graphene. Science. 2010;327(5966):662-.
[13]    Kim KS, Zhao Y, Jang H, Lee SY, Kim JM, Kim KS, et al. Large-scale pattern growth of graphene films for stretchable transparent electrodes. Nature. 2009;457(7230):706-10.
[14]    Pereira VM, Castro Neto AH. Strain engineering of graphene's electronic structure. Physical review letters. 2009;103(4):046801.
[15]    Gui GP, Kadayaprath G, Tan SM, Faliakou EC, Choy C, Ward A, et al. Long-term quality-of-life assessment following one-stage immediate breast reconstruction using biodimensional expander implants: the patient's perspective. Plast Reconstr Surg. 2008;121(1):17-24.
[16]    Guinea F, Katsnelson MI, Geim AK. Energy gaps and a zero-field quantum Hall effect in graphene by strain engineering. Nat Phys. 2010;6(1):30-3.
[17]    Bao WZ, Miao F, Chen Z, Zhang H, Jang WY, Dames C, et al. Controlled ripple texturing of suspended graphene and ultrathin graphite membranes. Nat Nanotechnol. 2009;4(9):562-6.
[18]    Low T, Guinea F. Strain-induced pseudomagnetic field for novel graphene electronics. Nano letters. 2010;10(9):3551-4.
[19]    Ni ZH, Wang HM, Ma Y, Kasim J, Wu YH, Shen ZX. Tunable stress and controlled thickness modification in graphene by annealing. Acs Nano. 2008;2(5):1033-9.
[20]    Bekyarova E, Itkis ME, Ramesh P, Berger C, Sprinkle M, de Heer WA, et al. Chemical Modification of Epitaxial Graphene: Spontaneous Grafting of Aryl Groups. J Am Chem Soc. 2009;131(4):1336-+.

[21]     Elias DC, Nair RR, Mohiuddin TMG, Morozov SV, Blake P, Halsall MP, et al. Control of Graphene's Properties by Reversible Hydrogenation: Evidence for Graphane. Science. 2009;323(5914):610-3.
[22]     Zhang H, Bekyarova E, Huang JW, Zhao Z, Bao WZ, Wang FL, et al. Aryl Functionalization as a Route to Band Gap Engineering in Single Layer Graphene Devices. Nano letters. 2011;11(10):4047-51.
[23]     Wang QH, Jin Z, Kim KK, Hilmer AJ, Paulus GLC, Shih CJ, et al. Understanding and controlling the substrate effect on graphene electron-transfer chemistry via reactivity imprint lithography. Nat Chem. 2012;4(9):724-32.
[24]     Childres I, Jauregui LA, Foxe M, Tian JF, Jalilian R, Jovanovic I, et al. Effect of electron-beam irradiation on graphene field effect devices. Appl Phys Lett. 2010;97(17).
[25]     Xu M, Fujita D, Hanagata N. Monitoring electron-beam irradiation effects on graphenes by temporal Auger electron spectroscopy. Nanotechnology. 2010;21(26):265705.
[26]     Teweldebrhan D, Balandin AA. Modification of graphene properties due to electron-beam irradiation. Appl Phys Lett. 2009;94(1).
[27]     Velasco J, Jr., Zhao Z, Zhang H, Wang F, Wang Z, Kratz P, et al. Suspension and measurement of graphene and Bi2Se3 thin crystals. Nanotechnology. 2011;22(28):285305.
[28]     Bolotin KI, Sikes KJ, Jiang Z, Klima M, Fudenberg G, Hone J, et al. Ultrahigh electron mobility in suspended graphene. Solid State Commun. 2008;146(9-10):351-5.
[29]     Du X, Skachko I, Barker A, Andrei EY. Approaching ballistic transport in suspended graphene. Nat Nanotechnol. 2008;3(8):491-5.
[30]     Zhang H, Bao W, Zhao Z, Huang J-W, Standley B, Liu G, et al. Visualizing Electrical Breakdown and ON/OFF States in Electrically Switchable Suspended Graphene Break Junctions. Nano letters. 2012;12(4):1772-5.
[31]     Cadelano E, Palla PL, Giordano S, Colombo L. Nonlinear Elasticity of Monolayer Graphene. Physical review letters. 2009;102(23).
[32]     Lee C, Wei X, Kysar JW, Hone J. Measurement of the elastic properties and intrinsic strength of monolayer graphene. Science. 2008;321(5887):385-8.
[33]     Cocco G, Cadelano E, Colombo L. Gap opening in graphene by shear strain. Physical Review B. 2010;81(24):241412.
[34]     Huang M, Pascal TA, Kim H, Goddard WA, Greer JR. Electronic−Mechanical Coupling in Graphene from in situ Nanoindentation Experiments and Multiscale Atomistic Simulations. Nano letters. 2011;11(3):1241-6.
[35]     Gui G, Li J, Zhong J. Band structure engineering of graphene by strain: First-principles calculations. Physical Review B. 2008;78(7):075435.
[36]     Brown L, Hovden R, Huang P, Wojcik M, Muller DA, Park J. Twinning and twisting of tri- and bilayer graphene. Nano letters. 2012;12(3):1609-15.

[37]     Ping J, Fuhrer MS. Layer number and stacking sequence imaging of few-layer graphene by transmission electron microscopy. Nano letters. 2012;12(9):4635-41.